\documentclass[twocolumn]{aastex62}
\hypersetup{linkcolor=blue,citecolor=blue,filecolor=cyan,urlcolor=magenta}

\accepted{October 27, 2018}

\submitjournal{ApJL}

\shorttitle{SO$_2$ in MonR2 IRS3}
\shortauthors{Dungee et al.}

\begin{document}

\title{High-Resolution SOFIA/EXES Spectroscopy of $\mathrm{SO}_2$ Gas in the Massive Young Stellar Object MonR2 IRS3: Implications for the Sulfur Budget}

%% AUTHORS

%% Include each other on their own "\author" command.
%% \author[xxxx-xxxx-xxxx-xxxx]{Author Name}
%% where the the "x" are the ORCID of the author. If you don't have it leave out the "[]" part of the command.

%% The new \correspondingauthor command is available in V6.2 to identify the
%% corresponding author of the manuscript. 
%% Use \email to set provide email addresses. Each \email will appear on its
%% own line so you can put multiple email address in one \email call. 

\correspondingauthor{Ryan Dungee}
\email{rdungee@hawaii.edu}

\author[0000-0001-6669-0217]{Ryan Dungee}
\affil{Institute for Astronomy, University of Hawaii, 2680 Woodlawn Dr, Honolulu, HI 96822, USA}

\author[0000-0001-9344-0096]{Adwin Boogert}
\affil{Institute for Astronomy, University of Hawaii, 2680 Woodlawn Dr, Honolulu, HI 96822, USA}

\author{Curtis N. DeWitt}
\affil{USRA, SOFIA, NASA Ames Research Center, MS 232-11, Moffett Field, CA 94035, USA}

\author{Edward Montiel}
\affil{Department of Physics, University of California Davis, 1 Shields Ave, Davis, CA 95616, USA}

\author[0000-0002-8594-2122]{Matthew J. Richter}
\affil{Department of Physics, University of California Davis, 1 Shields Ave, Davis, CA 95616, USA}

\author[0000-0003-4909-2770]{Andrew G. Barr}
\affil{Leiden Observatory, Leiden University, PO Box 9513, 2300 RA, Leiden, The Netherlands}

\author{Geoffrey A. Blake}
\affil{Division of Geological and Planetary Sciences, MC 150-21, California Institute of Technology \\ 1200 E California Blvd., Pasadena, CA 91125, USA}
\affil{Division of Chemistry and Chemical Engineering, California Institute of Technology \\ 1200 E California Blvd., Pasadena, CA 91125, USA}

\author{Steven B. Charnley}
\affil{NASA Goddard Space Flight Center, 8800 Greenbelt Road, MD 20771, USA}

\author{Nick Indriolo}
\affil{Space Telescope Science Institute, 3700 San Martin Drive, Baltimore, MD 21218, USA}

\author{Agata Karska}
\affil{Centre for Astronomy, Faculty of Physics, Astronomy and Informatics \\ Nicolaus Copernicus University, Grudziadzka 5, 87-100 Torun, Poland}

\author{David A. Neufeld}
\affil{Department of Physics and Astronomy, Johns Hopkins University, 3400 N. Charles St, Baltimore, MD 21218, USA}

\author{Rachel L. Smith}
\affil{North Carolina Museum of Natural Sciences, 121 West Jones St, Raleigh, NC 27603, USA}
\affil{Department of Physics and Astronomy, Appalachian State University, 525 Rivers St, Boone, NC 28608-2106, USA}

\author{Alexander G. G. M. Tielens}
\affil{Leiden Observatory, Leiden University, PO Box 9513, 2300 RA, Leiden, The Netherlands}

\begin{abstract}

Sulfur has been observed to be severely depleted in dense clouds leading to uncertainty in the molecules that contain it and the chemistry behind their evolution.
Here, we aim to shed light on the sulfur chemistry in young stellar objects (YSOs) by using high-resolution infrared spectroscopy of absorption by the $\nu_3$ rovibrational band of SO$_2$ obtained with the Echelon-Cross-Echelle Spectrograph on the Stratospheric Observatory for Infrared Astronomy.
Using local thermodynamic equilibrium models we derive physical parameters for the SO$_2$ gas in the massive YSO MonR2 IRS3.
This yields a SO$_2$/$\mathrm{H}$ abundance lower limit of $5.6\pm0.5\times10^{-7}$, or $>\!4\%$ of the cosmic sulfur budget, and an intrinsic line width (Doppler parameter) of $b<3.20\;\mathrm{km\;s}^{-1}$.
The small line widths and high temperature ($T_\mathrm{ex}=234\pm15\;\mathrm{K}$) locate the gas in a relatively quiescent region near the YSO, presumably in the hot core where ices have evaporated.
This sublimation unlocks a volatile sulfur reservoir (e.g., sulfur allotropes as detected abundantly in comet 67P/Churyumov--Gerasimenko), which is followed by SO$_2$ formation by warm, dense gas-phase chemistry.
The narrowness of the lines makes formation of SO$_2$ from sulfur sputtered off grains in shocks less likely toward MonR2 IRS3.

\end{abstract}


%% and \citet commands to identify citations.  

\section{Introduction} \label{sec:intro}

\begin{deluxetable*}{rccccc}[t]
\tablecaption{Observation log \label{tab:obslog}}
\tablecolumns{6}
\tablenum{1}
\tablewidth{0pt}
\tablehead{
\colhead{Target} &
\colhead{UTC start time} &
\colhead{Altitude} &
\colhead{Latitude} & \colhead{Longitude} & \colhead{Elevation} \\
\colhead{} & \colhead{(YYYY-mm-dd hh:mm)} &
\colhead{(start/end)} & \colhead{(start/end)} & \colhead{(start/end)} & \colhead{(start/end)}
}
\startdata
MonR2 IRS3 & 2017-01-24 03:06 & 41000/42000 ft & $48.1^\circ$/$44.6^\circ$ N &
$98.7^\circ$/$113.8^\circ$ W & $33^\circ$/$37^\circ$ \\
Sirius & 2017-01-24 04:38 & 43000 ft & $44.3^\circ$/$41.5^\circ$ N &
$115.4^\circ$/$124.4^\circ$ W & $26^\circ$/$29^\circ$ \\
\enddata
\end{deluxetable*}

As a dense cloud begins to collapse into a star it reaches densities high enough (i.e. $n\gtrsim\!10^3\;\mathrm{cm^{-3}}$) to enable the formation of a variety of molecules, particularly in the icy mantles that form around dust grains.
Understanding the chemistry from which these molecules originate can provide insight into the processes by which stars and planets form.
Various molecules have been proposed as tracers of evolution in protostellar environments \citep{hatchell98, bucklefuller03}.
Furthermore, the molecules produced inside these dense clouds will become components of the comets and planetesimals that are created and thus enrich the planetary system that forms \citep{visser09}.

Sulfur is the tenth most abundant element in the universe and has a very rich chemistry, meaning that it is well suited for understanding these processes \citep{charnley97, hatchell98, bucklefuller03}.
In the solar system, sulfur is well studied in cometary bodies \citep{bockelee, calmonte} allowing us to use sulfur-bearing molecules to study the link between the dense cloud, protostellar envelope, and primitive solar system objects. Moreover, there is evidence to suggest that sulfur is necessary for life as we know it \citep{life}.

However, sulfur has long been measured to be significantly depleted in dense clouds relative to abundances measured in diffuse clouds, \ion{H}{2} regions, and the solar photosphere \citep{tieftrunk}.
While this depletion is true for several elements key to astrochemistry, it is especially true for sulfur which has been observed to have abundances in dense clouds as low as $5\%$ of the measured cosmic abundance \citep[and references therein]{boogert15icyuniverse}.
This depletion stands in spite of the variety of sulfur-bearing species that have already been observed in the gas phase in dense clouds and star forming regions by their rotational line emission \citep[e.g.,][]{obs1, hatchell98, vandertak, obs5, drozdovskaya}.
Ice-phase observations have proven particularly difficult with only the detection of $\mathrm{OCS}$ \citep{palumbo} and, tentatively, SO$_2$ \citep{boogert97, zasowski}.
Thus, the majority of the sulfur is either contained in refractory material \citep[e.g.,~FeS;][]{fes} or in alternate volatile molecules.

Here, we use high-resolution ($R=\lambda/\Delta\lambda=55,000$) mid-infrared spectra to further study gas-phase SO$_2$ molecules.
Mid-infrared wavelengths enable studying the SO$_2$ nearest the hot core of MonR2 IRS3 through its absorption of the warm dust continuum.
Previously, the Infrared Space Observatory (ISO) measured the absorption of SO$_2$ in this and several other massive young stellar objects \citep[YSOs;][]{keane}.
However, it was impossible to resolve individual lines at the resolution ($R=2000$) of ISO.
With the high-resolution Echelon-Cross-Echelle Spectrograph \citep[EXES;][]{richter} on the Stratospheric Observatory for Infrared Astronomy \citep[SOFIA;][]{temi} we can now measure the line width and investigate the location and chemical origin of this gas. For example, lines that are tens of km s$^{-1}$ wide would indicate shocks capable of sputtering sulfur off refractory grains \citep[][]{may}, while narrower lines in a warm gas would be a signature of ice sublimation by stellar heat.

\section{Observations and Data Reduction} \label{sec:obs}

\begin{figure*}[t]
\plotone{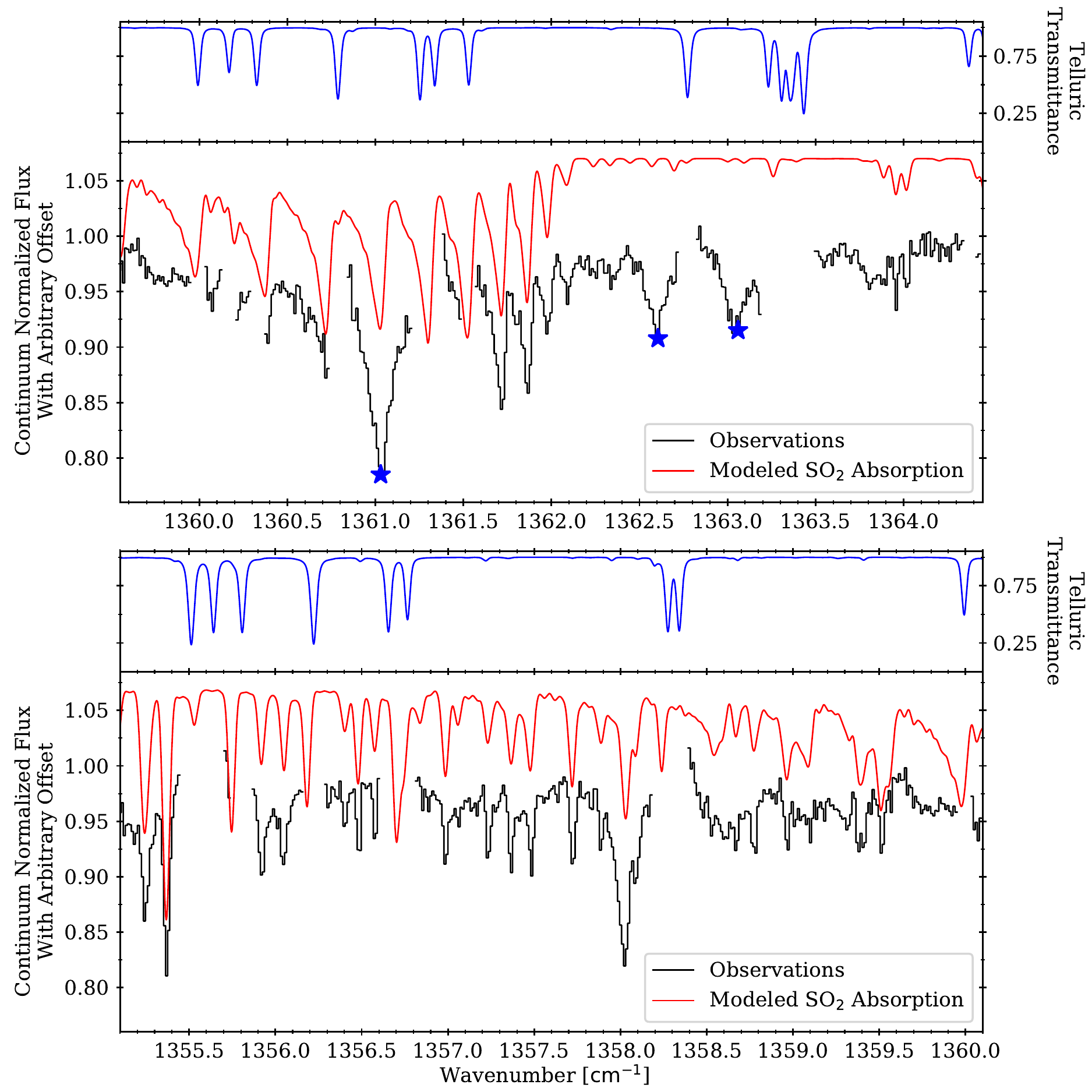}
\epsscale{0.8}
\caption{ \label{fig:fittodata} 
Subsets of the observed spectra of MonR2 IRS3 with the best-fit local thermodynamic equilibrium (LTE) model plotted in red; telluric transmission is plotted in blue.
Data are plotted in the rest frame of MonR2 IRS3 ($V_\mathrm{LSR}=10\;\mathrm{km\;s}^{-1}$).
For the best-fit parameters and their definitions, see Sec.~\ref{subsec:modeling}.
We use the upper limit Doppler parameter ($3.20\;\mathrm{km s^{-1}}$) in generating the plotted model.
Gaps in the data represent regions where the telluric transmission drops below $80\%$.
Blue stars denote absorption features due H$_2$O in the target.}
\end{figure*}

For the SO$_2$ observations (Table \ref{tab:obslog}), EXES was operated in the high-resolution configuration with a slit width of 3.2\arcsec{} providing for a spectral resolution ($R$) of $55,000\pm1100$ ($1\sigma$).
The resolution is assumed to be constant as a function of wavelength ($\lambda$) for a given slit width, and is extrapolated from $\mathrm{C_2H_2}$ gas cell absorption measurements at $\lambda=7.30\;\mathrm{\mu m}$.
The medium-resolution cross disperser was used with a slit length of 7.7\arcsec{}.
The spectra collected span from $7.23$ to $7.30$ and $7.31$ to $7.38\;\mathrm{\mu m}$, covering the $\nu_3$ rovibrational band of SO$_2$.

Data were reduced using the EXES instrument pipeline \citep[Redux;][]{pipeline} up until the order-merging step at which point our custom software was used.
First, we applied extra cuts to the data where the pipeline's reported signal-to-noise ratio (S/N) began decreasing ($\mathrm{S/N}<6.0$) at the edges of the instrument's blaze function.
Next, we applied telluric absorption corrections using atmospheric spectra generated by the Planetary Spectrum Generator \citep[PSG;][]{psg}.
The telluric lines were also used to improve the wavenumber calibration of the data to an accuracy of 0.005 cm$^{-1}$ (1 km s$^{-1}$), since the absorption features have wavenumbers known to high precision in the high-resolution transmission molecular absorption database \citep[HITRAN;][]{hitran}.
Subsequently, we divided our data by spectra of the standard star, Sirius, observed on the same flight to correct for fringes in the data.
Spectral orders were then stitched together by a combined process of linear interpolation and weighted averaging of the overlapping sections.
Finally, we normalized the spectra to the background continuum defined by a low amplitude, slowly varying sin function, fit to regions of the data least affected by absorption lines. A systematic continuum uncertainty of $\sim\!3\%$ is folded into the absorption line depth uncertainties.

MonR2 IRS3 was also observed with the NIRSPEC spectrometer \citep{mclean} at the Keck II telescope in the atmospheric $M$ band at $R=25,000$, as part of a survey exploring CO isotopologue abundances in a range of YSOs (R.~L.~Smith et al.~in prep.).
Here, the data for MonR2 IRS3 were used to measure gas-phase CO column densities and line profiles to determine SO$_2$/CO and, subsequently, SO$_2$/H abundance ratios (Sec.~\ref{subsec:abundances}).
For further information on these observations and their reduction see \citet{cocollab}.

\section{Modeling Absorption} \label{sec:model}

\begin{figure*}[t]
\plotone{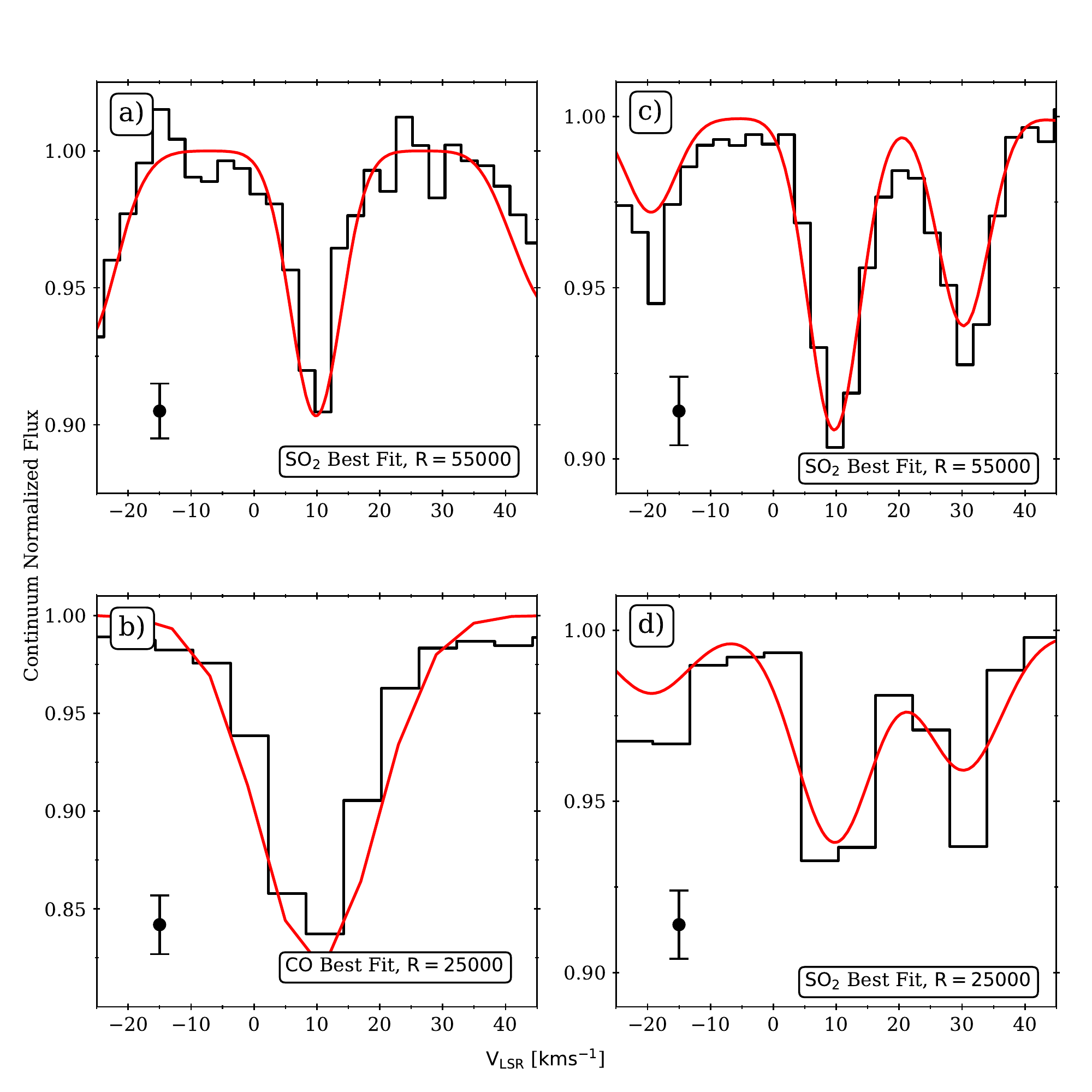}
\epsscale{0.8}
\caption{ \label{fig:profiles}
Isolated absorption features from each of our datasets showing line width comparisons.
The best-fit model is plotted in red. Note that the $b$=3.20 km/s used for SO$_2$ here is an upper limit for the intrinsic line width.
For the best-fit parameters and their definitions, see Sec.~\ref{subsec:modeling}.
Panels \textbf{a)} and  \textbf{b)} are isolated absorption features from the SO$_2$ and $^{13}$CO spectra, respectively.
Panels \textbf{c)} and \textbf{d)} are of the same SO$_2$ absorption feature, one convolved to the same resolution as our $^{13}$CO data.
The point with error bars represents the typical uncertainty for the observed spectra in that panel.} 
\end{figure*}

The observed spectra show many absorption features (Fig.~\ref{fig:fittodata}).
We used local thermodynamic equilibrium (LTE) models to generate model absorption spectra from which we derive physical parameters.

\pagebreak

\subsection{Generating Model Spectra} \label{subsec:modeling}

The models assume SO$_2$ and $^{13}$CO are present in a uniform slab perpendicular to and covering the line of sight.
Given a column density ($N$, $\mathrm{cm}^{-2}$), an excitation temperature ($T_\mathrm{ex}$, $\mathrm{K}$), an intrinsic line width (or Doppler parameter, $b$, $\mathrm{km\;s}^{-1}$, related to the full-width at half maximum by $\mathrm{FWHM}=2b\sqrt{\ln{2}}$), and a Gaussian instrumental line spread function, the LTE model generates an absorption spectrum.

The molecular line parameters (Einstein $A$ coefficient, partition function table, and quantum numbers) were retrieved from the HITRAN database.
These parameters were used to calculate the population in each energy level for a gas with temperature $T_\mathrm{ex}$ and total column $N$ following the standard Boltzmann equation.
Subsequently, line equivalent widths were calculated and then converted to line profiles of width $b$ at infinite resolution.
Each line has a Voigt profile that is then convolved with a Gaussian profile to the instrumental resolution and used to compute a continuum normalized intensity.

\subsection{Fitting to the Data} \label{subsec:fitting}

The LTE model was fit across the whole spectrum simultaneously and the best fit was found by using a $\chi^2$ value as our maximum likelihood estimator.
To obtain the uncertainties, we used a Markov Chain Monte Carlo (MCMC) sampling to determine the posterior distributions of our model parameters.
From these posterior distributions we obtained the 68\% credibility interval.
To do the sampling we used the open source Python package \texttt{emcee} \citep{emcee}.
Fig.~\ref{fig:fittodata} shows selected subsets of our best fits plotted over the data.
The Doppler shift of the target is not a parameter we fit for; instead, we adopted a value based off the $V_\mathrm{LSR}$ of $10\;\mathrm{km\;s}^{-1}$ measured in previous submillimeter studies of MonR2 IRS3 \citep{vandertak}, and in good agreement with our SO$_2$ and $^{13}$CO observations (Fig.~\ref{fig:profiles}).

Concurring with previous studies \citep{giannakopoulou, cocollab},  we fit two temperature components to the observations for both SO$_2$ and $^{13}$CO.
The priors common to all of our fits were the restrictions that $T_\mathrm{ex} > 1$, $N > 0.0$, and $b > 0.0$. We also folded an uncertainty on spectral resolution into our determination of $b$, the prior for this was a Gaussian distribution.
For EXES the peak probability occurred at $R=55,000$ with a standard deviation of $1100$, and for NIRSPEC the peak probability occurred at $R=25,000$ and was uncertain to 10\% at the $3\sigma$ level.

Fitting for the warm SO$_2$ alone, we found a $T_\mathrm{ex}$ of $234\pm15\;\mathrm{K}$, an $N$ of $4.95^{+0.29}_{-0.30}\times10^{16}\;\mathrm{cm}^{-2}$, and an upper limit of $b<3.20\;\mathrm{km\;s}^{-1}$.
The cold component was much less well determined, and so we only quote upper limits at the 95\% confidence level: $T_\mathrm{ex} < 88\;\mathrm{K}$ and $N < 1.3\times10^{14}\;\mathrm{cm}^{-2}$, assuming the same upper limit on the Doppler parameter.

For the $^{13}$CO we first measured the Doppler parameter by stacking the absorption features in our spectrum and measuring the line width assuming a Gaussian line profile and $R$ of $25,000$ (this approach is not possible for the crowded SO$_2$ spectrum).
This yielded $b$ values of $7.4\pm2.2$ and $5.3\pm2.1\;\mathrm{km\;s}^{-1}$ for the warm and cold $^{13}$CO components, respectively.
These values were then used as additional priors.
The other key difference from our SO$_2$ analysis was that we fit the warm $^{13}$CO to the high-$J$ level transitions, where it is the only contributor, before fitting a cold component on top of the now determined warm component in the low-$J$ level transitions.
A single temperature component was unable to produce deep absorption features in both low-$J$ level transitions and high-$J$ level transitions.
This yielded a best fit for the warm $^{13}$CO component of $T_\mathrm{ex}=240\pm25\;\mathrm{K}$ and $N=1.1\pm0.2\times10^{17}\;\mathrm{cm}^{-2}$, and a cold component with $T_\mathrm{ex}=10\pm7\;\mathrm{K}$ and $N=3.7^{+0.6}_{-1.0}\times10^{16}\;\mathrm{cm}^{-2}$.
These values are consistent with those found by a curve of growth analysis \citep[R.~L.][2018, in preparation]{cocollab}.

\subsection{Abundances} \label{subsec:abundances}

Since the measured line width for the warm $^{13}$CO gas ($7.4\pm2.2$ km s$^{-1}$) is broader than that of the SO$_2$ gas ($<3.20$ km s$^{-1}$), we are possibly including additional gas not associated with the reservoir of SO$_2$ we observed.
Assuming a typical $^{12}$CO/$^{13}$CO = 80 and $\mathrm{H_2}$/$^{12}$CO = 5000 \citep{lacy}, we derive a lower limit on the abundance of SO$_2$ relative to $N_{\rm H}$ (=$N$(H)+2$N$(H$_2$)) of $(5.6\pm0.5)\times10^{-7}$ for the warm SO$_2$.
Comparing this lower limit to the cosmic sulfur abundance \citep[S/H=$1.3\times10^{-5}$,][]{cosmicsulfur} shows that this SO$_2$ gas accounts for $>4\%$ of the sulfur budget.
We place an upper limit on the cold SO$_2$ abundance of $4.4\times10^{-9}$,
by using a cold $^{12}$CO gas column of $N=80\times3.7\times10^{16}=3.0\times10^{18}\;\mathrm{cm}^{-2}$.
Frozen CO contributes little.
Using data from \citet{gibb} we derived a $^{12}$CO ice column upper limit of $0.5\times10^{17}\;\mathrm{cm}^{-2}$.

We also calculate an abundance relative to H$_2$O, allowing for direct comparison with cometary results.
\citet{boonman} reported a column density $N_{\mathrm{H_2O}}=5\pm2\times10^{17}\;\mathrm{cm^{-2}}$ with $250^{+200}_{-100}\;\mathrm{K}$.
This yields a warm abundance, SO$_2$/H$_2$O, of $>10\pm3\%$. Lacking measurements of the foreground H$_2$O, we could not do the same for our cold SO$_2$ measurements.
All abundances are summarized in Table \ref{tab:summary}.

\subsection{Comparison with Previous Work} \label{subsec:prevwork}
Millimeter-wave observations of MonR2 IRS3 have measured the SO$_2$ column density at beam sizes of $\sim\!15\arcsec{}$, probing the cool envelope \citep{vandertak}. The column density of $1.5\pm 0.3\times10^{14}\;\mathrm{cm}^{-2}$ is consistent with our cold SO$_2$ component's upper bound of $1.3\times10^{14}\;\mathrm{cm}^{-2}$. Additionally, our measured warm SO$_2$ component is consistent with infrared measurements by \citet{keane}, who reported $T_\mathrm{ex}=225^{+50}_{-70}$ and $N=4.0\pm0.8\times10^{16}\;\mathrm{cm}^{-2}$ using an adopted $b$ of $3\;\mathrm{km\;s^{-1}}$.

\section{Discussion} \label{sec:disc}

\begin{deluxetable*}{lcc|c|cc|c}[t]
\tablecaption{Abundances of sulfur-bearing molecules toward MonR2 IRS3 and Comet 67/P. \label{tab:summary}}
\tablecolumns{11}
\tablenum{2}
\tablewidth{0pt}
\tablehead{
\colhead{Species} &
\multicolumn{2}{c}{Hot Core\tablenotemark{a}} &
\multicolumn{1}{c}{Foreground Gas\tablenotemark{b}} &
\multicolumn{2}{c}{Foreground Ice} &
\colhead{Comet 67/P's Coma} \\
\colhead{} &
\colhead{$X_{\mathrm{H}}$} & \colhead{$X_{\mathrm{H_2O}}$} & 
\colhead{$X_{\mathrm{H}}$} & 
\colhead{$X_{\mathrm{H}}$\tablenotemark{c}} & \colhead{$X_{\mathrm{H_2O}}$\tablenotemark{d}} &
\colhead{$X_{\mathrm{H_2O}}$} \\
\colhead{} &
\colhead{$10^{-7}$} & \colhead{\%} &
\colhead{$10^{-7}$} &
\colhead{$10^{-7}$} & \colhead{\%} &
\colhead{\%}
}
\startdata
SO$_2$ & $>5.6\pm0.5$ & $10\pm3$ & $<0.044$ & $<5.7$ (1) & $<0.6$ (1) & $0.127\pm0.003$ (4) \\
$\mathrm{H_2S}$ & - & - & - & $<2.8$ (2) & $<1.1$ (2) & $1.10\pm0.05$ (4) \\
$\mathrm{OCS}$ & - & - & - & $<0.18$ (3) & $<0.07$ (3) & $0.041\pm0.001$ (4) \\
$\mathrm{S_2}$ & - & - & - & - & - & $0.197\pm0.003$ (4) \\
\enddata
\tablenotetext{a}{Calculated from an SO$_2$ column of $4.95\times10^{16}\;\mathrm{cm^{-2}}$, and a derived hydrogen column of $8.8\times10^{22}\;\mathrm{cm^{-2}}$ or an H$_2$O column of $5\times10^{17}\;\mathrm{cm^{-2}}$ (Sec.~\ref{subsec:abundances}).}
\tablenotetext{b}{Calculated from an SO$_2$ column of $1.3\times10^{14}\;\mathrm{cm^{-2}}$, and a hydrogen column of $3.0\times10^{22}\;\mathrm{cm^{-2}}$ (Sec.~\ref{subsec:abundances}).}
\tablenotetext{c}{Relative to a hydrogen column of $3.0\times10^{22}\;\mathrm{cm^{-2}}$ derived from our cold $^{13}$CO gas column (Sec.~\ref{subsec:abundances}).}
\tablenotetext{d}{Relative to an ice column $N_\mathrm{H_2O}=1.9\times10^{18}\;\mathrm{cm^{-2}}$ \citep{gibb}.}
\tablecomments{Sources: (1) derived from data in \citet{gibb}, (2) \citet{smith}, (3) \citet{palumbo}, (4) \citet{calmonte}.}
\end{deluxetable*}

The abundance of warm SO$_2$ gas is over two orders of magnitude higher than the cold gas, suggesting a sulfur reservoir that is unlocked after heating.
The small line widths ($b<3.20\;\mathrm{km\;s}^{-1}$) likely imply a yet unobserved precursor in the ice, rather than in the refractory materials.
Moreover, the warm gas-phase SO$_2$/H$_2$O ratio of $10\pm3\%$ is at least a factor 10 larger than that observed in the foreground ice toward this target (Table \ref{tab:summary}).
All of this hints at an efficient gas-phase process that converts the sublimated sulfur-bearing ice molecules into SO$_2$.

\subsection{SO$_2$ Formation} \label{subsec:narrowlines}

Shocks have previously been observed to lead to substantial enhancements in SO$_2$ abundances \citep{pineau, so2shocks, hh212so2}.
The extreme temperatures enable a variety of gas-phase reactions of pre-shock gas or sublimated ices that enhance the formation of SO$_2$. 
Indeed, shock chemistry models successfully replicated the measured abundance of gas-phase SO$_2$ toward the Orion Plateau, which shows very broad lines indicative of shocks generated by Orion IRc 2 outflows \citep[$b\gtrsim 12-15\;\mathrm{km\;s}^{-1}$, ][]{blake}.
Also, shocks of tens of km s$^{-1}$ can shatter or sputter dust grains \citep{may}, possibly leading to the release of more sulfur and subsequent SO$_2$ formation.
However, the SO$_2$ lines toward MonR2 IRS3 are substantially narrower ($b<3.20\;\mathrm{km\;s}^{-1}$; Fig.~\ref{fig:profiles}) than those in the Orion Plateau.
Our results are therefore more consistent with release from the ices due to radiative heating (or perhaps mild shocks) rather than from refractory grains in strong shocks.
Also, the SO$_2$/$\mathrm{H}$ abundance derived for MonR2 IRS3 ($0.6\times10^{-6}$) is somewhat higher than that measured in Orion IRc 2 \citep[$0.2\times10^{-6}$,][]{blake} and consistent with the range reported for the outflow target HH 212 \citep[$(0.4-1.2)\times10^{-6}$,][]{hh212so2}.
The formation of  SO$_2$ in hot cores is thus at least as efficient as in shocks and, importantly, sputtering of sulfur off of grains does not seem to be a required process.

This process, for the first time observed in a massive hot core, might also be important in lower-mass YSOs.
The strong enhancement of SO molecules observed in a protoplanetary disk \citep{booth} might thus relate to ice sublimation rather than grain sputtering in shocks.

\subsection{Progenitor Species} \label{subsec:progen}
There still remains the question of which molecular species lead to the efficient formation of gas-phase SO$_2$.
The large mismatch between ice-phase detections of SO$_2$ \citep{boogert97, zasowski} and the warm component we measured indicate that SO$_2$ cannot be sublimating directly from the ice.
Chemical models generally rely on large abundances of sublimated $\mathrm{H_2S}$ ice for SO$_2$ formation \citep{charnley97, woods}.
Indeed, measurements by the ROSINA experiment have shown that $\mathrm{H_2S}$ is the largest contributor to the sulfur budget in the comet 67P/Churyumov--Gerasimenko \citep{calmonte}.
In stark contrast, $\mathrm{H_2S}$ ice measurements toward dense clouds and protostellar envelopes yielded upper limits \citep{smith, jimenez} a factor of two below our gas-phase SO$_2$ abundance (Table \ref{tab:summary}), and we thus hesitate to conclude this is the primary source of sulfur that is driven into SO$_2$ molecules.
Instead, we consider the release of sulfur allotropes (e.g., $\mathrm{S_2}$, $\mathrm{S_3}$, $\mathrm{S_4}$) that have been shown to be the second most abundant sulfur carrier in the ices of comet 67P \citep{calmonte,   drozdovskaya}. These allotropes can be broken apart by helium atoms, allowing for gas-phase reactions between $\mathrm{OH}$, $\mathrm{O_2}$, and $\mathrm{S}$ that lead to the formation of SO$_2$ \citep{heliumatoms}.

Chemical models suggest that at temperatures $\gtrsim\!230\;\mathrm{K}$ the oxygen is preferentially driven into H$_2$O \citep{charnley97, doty, wishprogram}.
The measured SO$_2$ temperature is comparable to this threshold. This suggests the gas-phase formation of SO$_2$ in MonR2 IRS3 is suppressed in favor of the production of H$_2$O, though our results are not inconsistent with sub-$230\;\mathrm{K}$ temperatures.
Alternatively, the SO$_2$ we observe formed before further heating of the gas, or at larger, cooler radii.
High H$_2$O abundances were measured along this line of sight, though there remains the possibility that these observations probed a warmer region closer to the hot core than the SO$_2$ gas we observed. This cannot be excluded following observations by \citet{boonman}, who measured a temperature of $250^{+200}_{-100}\;\mathrm{K}$.
A high spectral resolution analysis by N.~Indriolo et al.~(in prep.) is expected to shed more light on this possibility.

\subsection{Critical Density} \label{subsec:critdens}
The basis for the LTE assumption is that collisional excitations at least match the rate of radiative de-excitations.
The minimum density at which this occurs is called the critical density, and values for SO$_2$ are of order $10^6\;\mathrm{cm^{-3}}$ to $10^7\;\mathrm{cm^{-3}}$ \citep{wakelam, williams}.
Using a modeled temperature profile for a hot core from \citet{vandertak99} and the radial density profile of MonR2 IRS3 reported by \citet{vandertak00}, we find that SO$_2$ with a temperature of $230\;\mathrm{K}$ is expected to reside at a radius of $\sim$500 au with a particle density of $n=6\times10^6\;\mathrm{cm^{-3}}$, comparable to the SO$_2$ critical densities.
Also, the critical density of $^{13}$CO is over an order of magnitude smaller than that of SO$_2$ and thus $^{13}$CO is most likely in LTE.
The similarity of the excitation temperatures of these molecules thus further indicates that the SO$_2$ transitions are thermalized.

\section{Conclusions} \label{sec:conc}
We have measured a warm SO$_2$ gas with temperature $234\pm15\;\mathrm{K}$ and an abundance with respect to hydrogen $>\!5.6\pm0.5\times10^{-7}$ with narrow line widths ($b<3.20\;\mathrm{km\;s}^{-1}$) in the massive YSO MonR2 IRS3.
These infrared absorption values confirm the existence of a large reservoir of sulfur that has been unobserved in past submillimeter observations.
Moreover, this warm SO$_2$ contributes $>\!4\%$ of the sulfur budget in our target.
The small line widths are inconsistent with SO$_2$ formation from sulfur sputtered off refractory grains in strong shocks.
Thus, we conclude the abundant SO$_2$ gas likely originates from sublimated ices in the hot core close to the massive YSO.
Past $\mathrm{H_2S}$ and SO$_2$ ice observations indicate that they are unlikely precursors to this gas, leading us to conclude there is a large reservoir of sulfuretted molecules in the ice that has yet to be observed.
Based on comet 67P measurements these might be sulfur allotropes.
Future observations of sulfur-containing ices in samples of dense clouds and YSOs by the planned {\it James Webb Space Telescope} should help us to confirm this hypothesis.
However, the direct detection of sulfur allotropes is unlikely, considering their infrared absorption bands are weak and broad \citep[17--50$\;\mathrm{\mu m}$; e.g.,][]{mahjoub}.
High spectral resolution observations of gas-phase SO$_2$ with EXES on SOFIA toward a larger sample of YSOs are also needed to further distinguish between quiescent hot core and shocked environments.
Additionally, higher-resolution spectra for $^{13}$CO obtained by an instrument such as iSHELL \citep{ishell} will allow for better characterization of the dynamical components from the $^{13}$CO line profile and SO$_2$ abundances in each component.

\acknowledgments

Based in part on observations made with the NASA/DLR Stratospheric Observatory for Infrared Astronomy (SOFIA). SOFIA is jointly operated by the Universities Space Research Association, Inc. (USRA), under NASA contract NNA17BF53C, and the Deutsches SOFIA Institut (DSI) under DLR contract 50 OK 0901 to the University of Stuttgart. Financial support for this work was provided by NASA through award No.~SOF 04-153 issued by USRA.

Some of the data presented herein were obtained at the W. M. Keck Observatory, which is operated as a scientific partnership among the California Institute of Technology, the University of California and the National Aeronautics and Space Administration. The Observatory was made possible by the generous financial support of the W. M. Keck Foundation. The authors wish to recognize and acknowledge the very significant cultural role and reverence that the summit of Maunakea has always had within the indigenous Hawaiian community. We are most fortunate to have the opportunity to conduct observations from this mountain.

A.~K.~acknowledges support from the Polish National Science Center grant 2016/21/D/ST9/01098.

R.~L.~S.~gratefully acknowledges support under NASA Emerging Worlds grant NNX17AE34G.

\vspace{5mm}
\facilities{SOFIA}
\software{Redux \citep{pipeline}, emcee \citep{emcee}, PSG \citep{psg}}

\end{document}